# Practical Location Validation in Participatory Sensing Through Mobile WiFi Hotspots

Francesco Restuccia, Andrea Saracino, and Fabio Martinelli

*Abstract*— The reliability of information in participatory sensing (PS) systems largely depends on the accuracy of the location of the participating users. However, existing PS applications are not able to efficiently validate the position of users in large-scale outdoor environments. In this paper, we present an efficient and scalable *Location Validation System* (LVS) to secure PS systems from location-spoofing attacks. In particular, the user location is verified with the help of mobile WiFi hot spots (MHSs), which are users activating the WiFi hotspot capability of their smartphones and accepting connections from nearby users, thereby validating their position inside the sensing area. The system also comprises a novel verification technique called *Chains of Sight*, which tackles collusion-based attacks effectively. LVS also includes a reputation-based algorithm that rules out sensing reports of location-spoofing users. The feasibility and efficiency of the WiFi-based approach of LVS is demonstrated by a set of indoor and outdoor experiments conducted using off-the-shelf smartphones, while the energy-efficiency of LVS is demonstrated by experiments using the *Power Monitor* energy tool. Finally, the security properties of LVS are analyzed by simulation experiments. Results indicate that the proposed LVS system is energy-efficient, applicable to most of the practical PS scenarios, and efficiently secures existing PS systems from location-spoofing attacks.

*Index Terms*—Participatory Sensing, Smartphones, Security, WiFi Hotspots, Location Spoofing.

## I. INTRODUCTION

Undoubtedly, smartphones have become one of the most powerful and pervasive technologies today. Among all features, the simplicity of use make smartphones ideally suited for a novel and tremendously potential sensing paradigm, known as *participatory sensing* (PS) [1]. The basic idea behind PS is to allow ordinary citizens to participate in large-scale sensing surveys with the help of user-friendly applications installed in their smartphones. This not only reduces dramatically deployment costs of fixed infrastructures, but also provides fine-grained spatio-temporal coverage of the sensing area. Significant research and development from industry and academia has been devoted to design PS systems improving life experience of users. Indeed an abundance of real-life applications, which take advantage of both low-level sensor data and high-level user activities, range from real-time traffic monitoring [2] [3] to air pollution or garbage monitoring [4]–[6]

F. Restuccia is with the Department of Electrical and Computer Engineering, Northeastern University, Boston, MA (e-mail: f.restuccia@northeastern.edu) ).

A. Saracino and F. Martinelli are with the Istituto di Informatica e Telematica del Consiglio Nazionale delle Ricerche, Via G. Moruzzi n.1, 56124, Pisa, Italy (e-mail: {a.saracino, f.martinelli}@iit.cnr.it).

to social networking [7], to name a few. For a complete survey of PS applications, the readers may refer to [8].

Above all features, the most innovative aspect of PS systems is that they are infrastructure-free, and rely only on the users' active participation to gather reliable data about the sensing area. Therefore, it becomes paramount to verify with relative precision the current users' *location*, since the sensed data (e.g., temperature) is significantly dependent on the spatial context. However, smartphone applications (apps) like `LocationHolic` or `FakeLocation` make extremely easy for users to spoof their current GPS location. Such software provides users with easy-to-use interfaces to manually set the global position system (GPS) coordinates, thereby deceiving location-based apps running on the device.

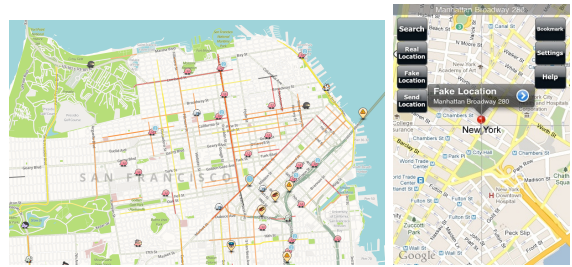

Fig. 1: Screenshot of Waze and `FakeLocation` apps.

### A. Motivation

Large-scale LSAs have potentially devastating consequences in terms of reliability and revenue loss of the PS systems [9]–[13]. To motivate and demonstrate the simplicity and impact of LSA, we considered the well-known traffic monitoring application `Waze`, which is a community-driven application gathering some complementary map data and other traffic information from users. Similar to other location-based apps, Waze learns from users' driving times to provide routing and real-time traffic updates. This application is free to download and use, and people can report accidents, traffic jams, speed and police traps, and can update roads, landmarks, house numbers, and so on. Figure 1 depicts a screenshot of Waze and `FakeLocation` apps. Waze has a point-based reputation system based on the frequency of traffic reports and miles driven[1].

---
[1] https://www.waze.com/wiki/Your_Rank_and_Points

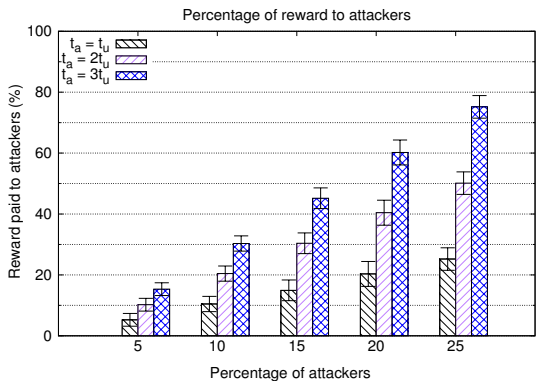

Fig. 2: Revenue given to attackers in [14].

In order to understand how much revenue the administration of the PS application may lose due to LSAs, we implemented and simulated the reward mechanism scheme due to Yang *et. al* (appeared in ACM MobiCom 2012 [14]). In particular, we focused our attention on the Platform-Centric reward model, in which the users are paid proportionally to the time $t_u$ they declare to dedicate to the sensing services (see [14] for additional details). We focused on this particular model because of its simplicity and its strong game-theoretical properties. Figure 2 shows the average percentage of revenue that attackers receive from the reward mechanism at each time step with respect to the total number of reward $R$ offered at each sensing request, as a function of the number of attackers and the declared time $t_a$ that will be dedicated to the sensing task (as multiple of $t_u = 1$ time unit). Figure 2 concludes that the amount of revenue the attackers steal from the reward mechanism grows linearly with the number of attackers.

In order to have an idea of the impact of attackers in term of revenue loss, let us suppose to have a percentage of attackers equal to 5%, 1000 users in the system, $R$ equal to $10, $t_a$ equal to $t_u$, and sensing requests every minute. Every time the PS application requests data from the users, the attackers steal $0.5 from the system, which means in a day the attackers steal $720. In a month and a year, respectively, the administration will lose $21,600 and $262,800, respectively. If the application requires data every 10 minutes, the attackers would still steal $26,280 a year.

*B. Our contribution*

The above examples demonstrate that LSAs tremendously undermine the reliability and the revenue of existing PS systems. However, given the extremely large scale of real-world PS systems and the uncontrollable, random mobility of smartphone users, verifying with relative accuracy the location of users becomes remarkably challenging. LSAs are also extremely difficult to detect, given the PS system has no means to find out whether users are using apps such as `FakeLocator`. This issue calls for a distributed solution which leverages the collective action of participating users.

This paper makes the following novel contributions.

- We propose the *Location Validation System* (LVS), which efficiently and effectively tackles LSA attacks. LVS exploits the collaborative actions of users and the WiFi capability of smartphones to validate the position of other users. In fact, two smartphones directly connected through WiFi range are practically sharing the same location, due to the limited WiFi range. In such way, these two users can mutually validate their locations inside the sensing area. By exploiting this technique on a large scale, LVS implements an effective, scalable and distributed anti location-spoofing system. A reputation-based algorithm is also proposed to filter out reports coming from malicious users.
- We propose the novel *Chain of Sights* (CoS) verification algorithm to contrast collusion-based attacks. CoS give a representation of the history of the validation process introduced through LVS. This process is based only on feedbacks sent by users to the system and does not rely on trusted control mechanisms. In fact, such mechanisms open the possibility of smart attacks based on colluding malicious users, which are part of a more challenging threat model not considered in former related work [15].
- The viability of the WiFi-based approach of LVS is demonstrated by real experiments conducted using off-the-shelf smartphones on indoor and outdoor testbeds. The results show that the framework is effective both in typical indoor environments, as well as in outdoor environments with greater distances between the users (up to 60m). We also measure the energy consumption of the WiFi-based mechanism of LVS using the Power Monitor [16] hardware tool. Results show that the proposed approach has practically no overhead on the smartphone resources in terms of energy consumption.
- The efficiency and effectiveness of the LVS framework against location-spoofing attacks is proven through simulations. Simulation results indicate that the framework is resilient to high percentages of attackers (up to 40%) and scales well with the number of users in the system.

This paper largely extends the work presented in [17] and [18], by introducing the concept of chain of sight, a deeper security analysis, an energy performance evaluation and formal demonstration of theorems.

The rest of the paper is organized as follows. Section II introduces preliminary concepts and formally defines the notion of LSA attack, while Section V discusses the related works. Section III describes in depth the proposed LVS framework, while Section IV presents experimental and simulation results of the LSA framework considering practical PS scenarios. Finally, Section VI draws conclusions with directions of future work.

## II. PRELIMINARIES

In this section, we first provide some technical background on PS and we define the threat model.

In this paper, we adopt the most common architecture for PS system, which is based on a PS platform (PSP) hidden inside a mobile cloud computing system [19]. Periodically, users are requested by the PSP to submit their sensed data to the PS system. At each request, users may choose to participate by sending their data to the PSP, or may simply ignore the request. The *sensing app*, which can be distributed through common application markets like *Google Play* or *App Store*, is responsible for providing the users a friendly interface for data visualization and acquisition, as well as ensuring reliable data communication between the users and PSP through cellular or WiFi network. After operations such as data filtering and aggregation, global information about the sensing area may be sent back to the users through the PS application, so as to be used for their daily activities.

As far as potential threats are concerned, we will assume that the communication between the users and the PS server runs is via reliable and protected wireless channel, where data cannot be lost, eavesdropped, modified or substituted. We also assume the PS server is totally reliable and trustworthy (root of trust), in particular, in terms of user registration, key management, issuing credentials, trust assessment and reputation management. Users are uniquely identified inside the network through an identifier (ID) which exploits a digest of the smartphone IMEI (International Mobile Equipment Identifier), which is unique for any device worldwide. Moreover, trying to modify a device IMEI (i.e., spoofing) is considered illegal and is extremely more complex than spoofing the location [20]. Therefore, given users cannot spoof their identity inside the systems, we assume that *sybil attacks are not possible*.

Henceforth, we will focus our attention to solving the attacks formalized in section III-E. In particular, attacks via the communication channels (e.g., eavesdropping, traffic jamming, etc.) are out of the scope of this paper.

## III. LOCATION VALIDATION SYSTEM

In this section we describe the LVS security framework to tackle the location-spoofing attack (LSA). We first describe the system model and formalize the LSA attack under such model. Next, we describe in details the algorithm used by LVS to select the users acting as mobile hot-spots (MHS), as well as the WiFi-based location validation algorithm of LVS. Finally, we describe the reputation-based algorithm used by LVS to filter out unreliable reports and therefore guarantee reliability of the PS system.

### A. System Model and Assumptions

Hereafter, we will suppose the smartphone sensing area is logically divided into $W$ *location areas* of size $S \times S$, in which $N$ users can move without restrictions (we do not assume any particular user mobility pattern and model). Specifically, users are free to move from one location area to another, and a given location area may contain any number of users (from 0 to $N$). However, users cannot be in two different location areas at the same time. We assume $S$ and $W$ are tuning parameters depending on the specific PS application and its required accuracy of user location. We assume the location area $L_k^t$ of user $u_k$ at time $t$ is identified by a pair of numerical coordinates representing a point in the two-dimensional Cartesian coordinate system $C \triangleq \{\mathcal{O}, \mathcal{X}, \mathcal{Y}\}$. We also assume that users are connected to the Internet through WiFi or 3/4G Internet connection.

- *LVS user module.* It is implemented inside the sensing app installed in the users' smartphones, and is responsible for handling the communication (through the Internet) between the PSP and the smartphone as far as the operations performed by LVS are concerned. It also handles the user WiFi hotspot activation as explained in III-C.
- *PSP communication module.* This module is implemented on the PSP, and is responsible for the communication (through the Internet) between the PSP and the LVS user module.
- *PSP computation module.* It handles the computation burden of LVS, which is the optimal selection of the users (see III-D) that will run the location validation algorithm (see III-C), as well as the computation of the Chain of Sights (see III-F) and the calculation of the reputation of the users (see III-G).

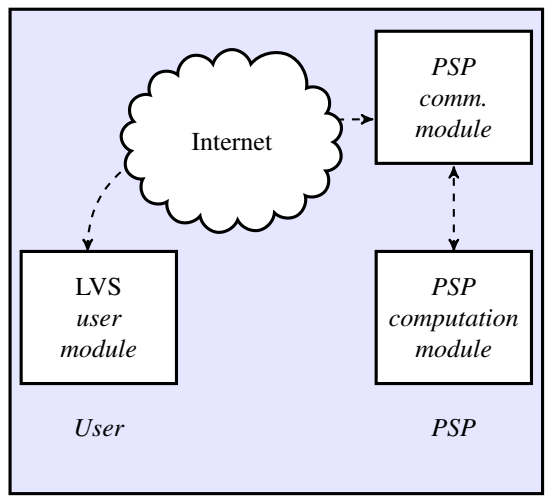

Fig. 3: Block diagram of LVS.

### B. Overview of LVS

Figure 3 depicts the four logical components of LVS. Let us describe the functionality of each module in detail.

- *LVS user module.* It is implemented inside the sensing app installed in the users' smartphones, and is responsible for handling the communication (through the Internet) between the PSP and the smartphone as far as the operations performed by LVS are concerned. It also handles the user WiFi hotspot activation as explained in III-C.
- *PSP communication module.* This module is implemented on the PSP, and is responsible for the communication (through the Internet) between the PSP and the LVS user module.
- *PSP computation module.* It handles the computation burden of LVS, which is the optimal selection of the users (see III-D) that will run the location validation algorithm (see III-C), as well as the computation of the Chain of Sights (see III-F) and the calculation of the reputation of the users (see III-G).

### C. Location Validation Algorithm

Before describing the LVS location validation algorithm, let us define as *mobile hot-spots* (MHSs) the subset of selected users (selection is explain in section III-D) who activate the built-in WiFi hotspot feature of their smartphones and wait for other users to connect. Users that reside in the WiFi range of MHSs are called *neighbors*. The neighbors and the MHS will mutually validate their locations.

The location validation algorithm divides time into validation *rounds*, occurring every $T_r$ time units; henceforth, we will refer to $t_j = j \cdot T_r$ as the time of the $j$-th validation round. During a validation round, the MHSs and their neighbors mutually validate their locations. A set of consecutive validation rounds is called validation *epoch* (Figure 4). The number of rounds composing a validation epoch and hence, its duration $T_e$, is variable and will be detailed later in the subsection.

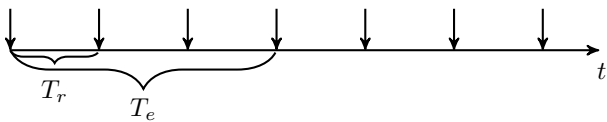

Fig. 4: Validation epoch timeline.

Let $N_i^j$ denote the number of users physically present in the $i$-th location area $i$ at time $t_j$. Also, let $D_i^{t_j}$ define the number of users advertising their position to be inside the location area $i$ at time $t_j$. During every validation round, the LVS validation algorithm performs the following three steps.

S1. The user module transmits her current location $L_k^{t_j}$ to the PSP computation module through the PSP communication module. For each location area, the PSP selects a subset of users among $D_i^{t_j}$ users that appear to be in the $i$-th location area (selection algorithm described in Section III-D).

S2. The selected users receive a message request from the PSP to act as MHS and validate the position of their neighbors through WiFi connection. At the same time, the neighbors also validate the position of the MHS for additional security. This is when the location validation takes place (details explained below), which we call *spotting* for brevity.

S3. Each user transmits the location validation information acquired in the current validation round to the PSP through the LVS user module. This information is used by the PSP computation module to compute users' reputation as detailed in sections III-F and III-G.

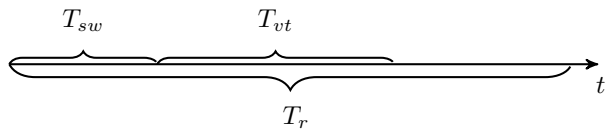

Fig. 5: Validation round timeline.

In detail, the operations performed by each user during each validation round in step S2 are summarized as follows.
- Each MHS turns on the WiFi hotspot capability, and after WiFi setup time $T_{sw}$ (see Figure 5), starts accepting connections from nearby users for a maximum validation time $T_{vt}$. Next,
    - each user $u_i$ connected to MHS $u_j$ sends a packet containing her unique ID number $ID_j$ to $u_j$ (remember that IDs cannot be spoofed);
    - MHS $u_j$ replies by sending a packet containing $ID_j$ to every neighbor;
    - after the reception of the packet from their MHS, the neighbors disconnect from their MHS.
- After the validation time $T_{vt}$ elapses, each MHS turns off the WiFi connection (if not active before the validation phase).
- Each user reports to the PSP the IDs of the users verified in the current validation round (if any).

This operation of mutual validation between an MHS and a user in its WiFi range is also called *spotting*. If the user $u_i$ is an MHS and the user $u_j$ is in its WiFi area, we say that $u_i$ spots $u_j$ and $u_j$ spots $u_i$. As an illustrative example, let us consider location area $A_i$ containing four users A, B, C and D at the validation round $j$ (Figure 6.a). During this round, the PSP chooses B to be MHS since she is close to users A and C. Users A and C are within the WiFi range of B, while D is in a different zone of $A_i$. Therefore, $A$ validates the location of $B$ and $C$, while both $B$ and $C$ validate the location of $A$. During round $j+1$ (Figure 6.b), $D$ and $B$ validate the location of each other.

As mentioned earlier, the PSP evaluates the reputation of each user once a validation epoch is finished. In particular, a *validation epoch* ends when the position of all users in a given location area has been validated by at least $q$ users, where $q$ is a system parameter. More formally, the duration of $j$-th validation epoch for location area $A_i$ is defined as $\min(e_M, e_{max})$, where $e_M$ is the number of validation rounds required to validate $M\%$ of the $D_i^{t_j}$ users by at least $q$ users, and $e_{max}$ is a system parameter.

Let us now discuss in details some aspects and advantanges of the location validation algorithm of LVS.

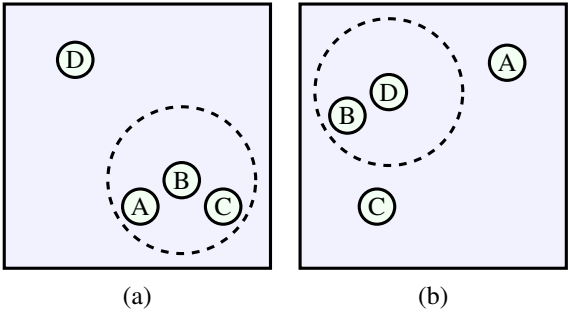

Fig. 6: Position of users at rounds $j$ and $j+1$.

- The operations performed by the LVS user module, included the activation of the WiFi hotspot capability, do *not* require manual activation by the user, but are instead handled automatically by the PSP through the user module of LVS. This allows the users to act as MHS without manual intervention, easing the burden on the users.
- Modern smartphones include functionalities enabling at the same time WiFi and 3/4G connection[2]. In particular, the functionality avoids disconnections if the user is using the WiFi interface for Internet connection when a validation round begins. Hence, before the user WiFi is disconnected to become an MHS or to connect to an MHS nearby, the connection is migrated to mobile data (3G/4G) to ensure continuity. This allows the location validation algorithm to run without disrupting existing connections on the user smartphone.
- The WiFi-based algorithm of LVS has the remarkable advantage that neighbors and MHS will *mutually* authenticate each other. This given additional security to the system, as more location validation information is available to the PSP to prevent LSAs.
- The cooperation of the user in acting as MHS can be guaranteed by using efficient and effective incentive mechanism, such as [14]. Therefore, the assumption that enough users acting as MHSs will be available at each validation round is sound.

### D. Selection of MHSs

The selection of MHSs at each validation round is an extremely important problem for the PS system performance. An MHS should be chosen in such a way that maximizes sensing area coverage and therefore, validates as many users as possible at each validation round. In this subsection, we define the problem of optimum selection of MHSs at each validation round and prove that it is an NP-Hard problem. In particular, we first select a minimum set of users to act as MHSs such that every other user with at least one neighbor may connect to at least one MHS. Next, we present an approximation algorithm which yields

[2]Available at https://play.google.com/store/apps/details?id=it.opbyte.superdownload

a suboptimal solution in polynomial time.

*Problem 1 (P1).* Let $U^j = \{u_1, ..., u_N\}$ the set of users of the PS system at time $t_j$ having at least one neighbor in the WiFi range. Given the position $L_k^j$, $1 \leq k \leq N$, of every user $u_k$ at time $t_j$, select a subset $P^j \subseteq U^j$ to act as MHSs such that (i) each user with at least one neighbor can connect to at least one MHS, and (ii) $|P^j|$ is minimum.

*Lemma 1:* Problem 1 is NP-Hard. Proof shown in the Appendix.

To solve P1, we use the greedy algorithm proposed by Chvatal in [21]. The complexity of the algorithm is $O(|P^j| \log |P^j|)$ with the best known sorting algorithm. It can be proven [21] that this algorithm achieves an approximation ratio of $H(|P^j|)$, where $H(n)$ is the $n$-th harmonic number, i.e., $H(n) = \sum_{k=1}^{n} \frac{1}{k} < \ln(n+1)$.

*Proof:* We prove the hardness of the problem by reducing the optimization version of the set cover problem (O-SCP), which is known to be NP-Hard, to P1. O-SCP is stated as follows. Given a set $X$ of elements (called the universe), and a set $S$ of $n$ subsets whose union equals the universe, identify the smallest subset of $S$ whose union equals the universe. By defining $V^j = X$, and $Z^j = X$, a set cover for $X$ is a solution $Q^j$ to P2, and viceversa. Therefore, O-SCP $\leq_P$ P1, which means P1 is NP-Hard. ∎

### E. Attacks Formalization

After defining the main components of LVS, it is possible to extend the threat model formally defining two attacks in addition to the LSA, formally redefined here for the sake of readability:

- *Location Spoofing Attack.* Let a PS system have $N$ active users $U^j = \{u_1, \ldots, u_N\}$ at time $t_j$ and $W$ location areas $\mathcal{A} = \{A_1, ..., A_W\}$. A location-spoofing attack (LSA) is performed when one or more users $u_s$ (called spoofers) belonging to the set $\mathcal{U}_s \subseteq U^j$ advertise to the PSP a position (*fake position*) in a location area $A_l^f$ (*fake location area*), while their real location is in the location area $A_l^r$ (*real location area*), where $A_l^f \neq A_l^r$. The location in $A_l^f$ is provided continuously by the spoofer. We assume that during the attack, spoofers can move from one location area into another, but the condition $A_l^f = A_l^r$ is never met.
- *Collusion Attack.* The collusion attack is performed when one or more sets of users $U_c = \{u_1, \ldots, u_c\}$ perform an LSA providing the location $A_k$ when they are in location areas different from $A_k$, and at each validation round each user in $U_c$ validate the fake position in $A_k$ of all other users in $U_c$. This attack can represent a situation in which a group of users is expected to be in a specific place, while they all are in different places. Thus, they collude mutually validating the fake position.
- *Fraud Covering Attack.* In the fraud covering attack a user $u_m$ performs an LSA providing a position in $A_j$,

but being located in $A_k$. At the same time, another user $u_f$ effectively residing in $A_j$ validates at each validation round the position of $u_m$ in $A_j$. With this attack, a user can be located in a low density area, in order not to be spotted by other nodes residing in $A_k$ and pretending to provide information on a different area.

*F. Chains of Sights*

Chains of Sight (CoS) represent the situation of a location area describing the series of direct and indirect spotting between users in a validation epoch. The CoSs have been designed to improve the performance of LVS and to effectively tackle the Collusion and Fraud Covering attack described formerly. During each validation epoch, each user keeps track of the users spotted in the various rounds and shares this knowledge with the users spotted in the following rounds. As an example, if the user $u_i$ spotted the user $u_j$ in the round $r$, when at the round $r+1$ is spotted by the user $u_k$, $u_i$ will tell to $u_k$ about the presence of $u_j$ in the area. Thus $u_k$ indirectly spots $u_j$ and also validates the $u_j$ position. This information is expressed through a CoS in the following form: $u_j \to u_i/u_k$, read as "$u_j$ sees $u_k$ through $u_i$". It stems that through CoSs it is possible to reduce the number of validation round per epoch. A CoS has two main elements: the spotted node, which is the user identifier on the right end of the chain and the *length* which is the number of users in the chain. Notice that the chain length cannot be greater than $\psi_{max}$. At the end of each validation round the collection of CoS stored by a user $u_i$ is defined as *user area knowledge* $\Omega_{u_i}$, while the collection of all user area knowledges compose the *global area knowledge*.

Formally the CoSs are generated according to the following algorithm;
1) At each sensing round, the MHSs send their current area knowledge to the spotted users.
2) Spotted users send their current area knowledge only to their hot spot.
3) Each user $U_l$, including hot spots, update its area knowledge $\Omega_l$ with the useful information from the received area knowledge(s), called $\Omega_r$, exploiting the following algorithm:

---
**Algorithm 1** Updating $\Omega_l$

  **for all** $\gamma_i$ in $\Omega_r$ **do**
    **if** $\gamma_i$ not in $\Omega_l$ **then**
      $\Omega_l = \Omega_l \cup \gamma_i$
    **end if**
  **end for**

---

In Figure 7 it is reported an example of validation epoch composed of four rounds in a specific location area, where a fraud covering attack is being performed. We will use this example to give a better understanding of the CoS algorithm. The user $C$ is located in a different location area and colludes with $B$ who will always validate the position of $C$ in the area of Figure 7. In the first validation round the users $B$ and $E$ are selected as MHS. Thus, $B$ validates the position of $D$ who is nearby and maliciously also validates the position of $C$, whilst $E$ validates the position of $F$. We recall that the authentication is mutual, thus $D$ and $C$ will also validate the location of $B$, while $F$ will validate the location of $E$. The area knowledge will be the following:

$$\Omega_A = \emptyset, \quad \Omega_B = \{B \to C; B \to D\}$$
$$\Omega_C = \{C \to B\}, \quad \Omega_D = \{D \to B\}$$
$$\Omega_E = \{E \to F\} \quad \Omega_F = \{F \to E\}$$

In the second round $B$ is spotted by the MHS $F$ and in the information exchange will tell that he spotted $D$ and $C$ in the former round. Thus, the knowledge of $F$ at the end of the second round will be:

$$\Omega_F = \{F \to E\ F \to D; F \to B, F \to B/C\}$$

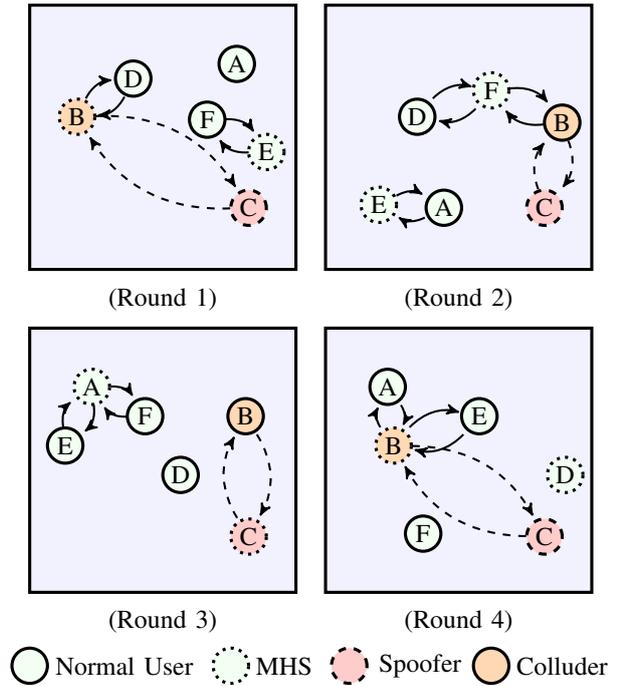

Fig. 7: Example of fraud covering.

For the sake of brevity we omit the knowledges of other users. In the third validation round, the user $C$ (spoofer) is selected as MHS together with $A$. $C$ validates the position of $B$ and no one else, since cannot see anyone in the area. In the forth round $B$ is chosen again as hot spot, together with $D$, and $B$ will validate again the position of $D$. The area knowledge of the various users at the end of round four is represented with a shortened notation in the following:

$$\Omega_A = \{A \to \{B, E, F\}; A \to B/D; A \to B/C\}$$

$$\Omega_B = \{B \to \{A, C, D, E, F\}\}$$
$$\Omega_C = \{C \to B; C \to B\{D, E, F\}\}$$
$$\Omega_D = \{D \to \{B, F\}; D \to F/E\}$$
$$\Omega_E = \{E \to \{A, B, F\}; E \to B/\{C, D\}\}$$
$$\Omega_F = \{F \to \{A, B, D, E\}, F \to B/C\}$$

At the end of the fourth validation round, each user has been spotted by at least four other users (80%). However, the presence of $C$ is validated directly only from $B$. Analyzing the CoSs it is easy to see that all other users validate the presence of $C$ indirectly, from the information received from $B$. It is unlikely that no other users directly validate $C$, thus, as formalized in the following, if this suspicious situation is repeated for a specific number of rounds, $C$ is considered a spoofer and $B$ a colluder.

Chains of sight are effective in contrasting the Collusion and Fraud Covering attack. In the collusion attack all the colluding users will provide to the system chains of sight where each user in the chain of sight belongs to the set $U_C$ of the colluding users. Assuming that the number of colluding users is lower than the average length of a CoS, for finding colluding users is sufficient to find CoSs that reports a set of users $U_S$ that are never spotted, directly or indirectly by users outside from this set. Formally, calling $\Omega_{u_k}$ the list of CoS owned by the user $u_k$ at a validation epoch, a collusion attack is detected if

$$\forall \{u_s \in U_S, u_k \in U_K\} \quad \Omega_{u_s} \cap \Omega_{u_k} = \emptyset$$

where $U_K$ is the set of all users declaring their position in the location area that are not part of $U_S$, i.e. $U_S \cap U_K = \emptyset$. In such a case, all users in $U_S$ are deemed as *malicious* after the threshold of $\theta_c$ validation epochs.

Chains of sight also allow to tackle the Fraud Covering Attack formerly described. In fact, the fake location of the user $u_m$ is only validated by the user $u_f$. Thus, every chain of sight that validates the position of $u_m$ will have the following two possible formats:

$$\begin{aligned} u_i &\to u_j / \ldots / u_f / u_m \\ u_f &\to u_m \end{aligned} \quad (1)$$

Given a threshold $\theta_f$, if for a number of validation epochs greater than $\theta_f$ all chains validating the position of $u_m$ are in the formats of Eq. 1, $u_m$ and $u_f$ are deemed as malicious.

### G. Reputation Algorithm

Let us now introduce the reputation model used by LVS to rule out the reports submitted by users spoofing their location. LVS assigns to each user $u_i$ reputation value $\rho_i^m$, which is updated at the end of the $m$-th validation epoch. In particular, the reputation $\rho_i^m$ of each user $u_i$ is updated after the end of the $m$-th validation epoch according to the following relation, inspired to the Jøsang reputation model [22]:

$$\rho_i^m = b_i^m - d_i^m - u_i^m$$

where $0 \leq \rho_i^m, b_i^m, d_i^m, u_i^m \leq 1$. In detail, $b_i^m$, $d_i^m$ and $u_i^m$ are respectively the *belief*, *disbelief* and *uncertainty* level associated to the reputation of user $u_i$ after the $m$-th validation epoch. These three values are updated at the end of the $m-1$-th validation epoch according to Algorithm 2.

---

**Algorithm 2** Updating $\rho_l$

$\rho_l = b_l - d_l - u_l$
$b_l + d_l + u_l = 1$
**for all** $u \in U^j$ **do**
  **if** $u$ location is verified **then**
    $b_l = b_l + \Delta_b$
    $u_l = u_l - \frac{\Delta_b}{2}$
    $d_l = d_l - \frac{\Delta_b}{2}$
  **else**
    **if** $u$ location is not verified **then**
      $u_l = u_l + \Delta_u$
      $b_l = b_l - \Delta_u$
    **end if**
  **else**
    **if** $u$ location is fake $\wedge$ $u$ is malicious **then**
      $d_l = d_l + \Delta_d$
      $b_l = b_l - \frac{\Delta_d}{2}$
      $u_l = u_l - \frac{\Delta_d}{2}$
    **end if**
  **end if**
**end for**

---

Let us now explain the algorithm in detail. By defining $A_i^d$ as the location area advertised by user $u_i$, the location of user $u_i$ is *verified* when at the end of a validation epoch her position has been validated by at least $q$ users. The location of user $u_i$ is *not verified* when, at the end of a validation epoch, less than $q$ users have validated the position of $u_i$ to be in the location area $A_l^d$. Finally, the location of $u_i$ is considered *fake* when her position has been validated by $q_e$ users in a location area $A_l^e \neq A_l^d$ and $q_e > q$. The user reputation is lowered as if her location is considered fake, when one or more of the conditions related to the collusion and fraud covering attacks applies.

Since the condition $b_l + d_l + u_l = 1$ must always hold, after each update the three components are normalized. We point out that $\Delta_b, \Delta_d$ and $\Delta_u$ are configurable parameters of the LVS framework and can be varied to best fit to different configurations with different values of $T_r$, $e_{max}$, user density and number of location areas.

## IV. RESULTS

The target of the experimental evaluation is to evaluate the viability of the WiFi-based approach of LVS in indoor and outdoor scenarios, as well as its energy-efficiency. In particular, in the following we validate the following assumptions.

- First, some time is necessary to ensure effective WiFi pairing between the MHS and neighbors. Moreover,

considering that smartphones are battery-powered devices, it is not sound to assume that devices will have their WiFi interfaces always active. Therefore, before the pairing phase, it will be necessary to wait until the WiFi interface becomes active. Such overhead must be reasonably low with respect to the total time the user and her MHS will be connected in the current validation round.

- Second, supposing that a normal moving user and a MHS are nearby when the validation protocol starts, if the user is not in the MHS range for enough time, the location validation will not occur.

To validate such assumptions, we performed experiments aimed at measuring the amount of successful mutual verification between two users. The experiments have been performed using two `Galaxy Nexus 4` with Android version 4.4. Experiments have been performed both indoor and outdoor, with different configurations, users' speeds, distances and movement patterns. Specifically, in each experiment the two users (hereafter referred to as U1 and U2) perform the following operations. At the beginning of each experiment, both U1 and U2 have their WiFi interface off. As soon as the experiment starts, U1 becomes an MHS and activates the built-in WiFi hot spot feature, while U2 simply turns on the WiFi interface and attemps to connect to U1. After 30 seconds from the beginning of each experiment (i.e., $T_{vt} = 30s$), U1 and U2 shut down their WiFi interfaces, ending the experiment. We developed a simple Android application which implements the LVS authentication protocol.

The indoor experiments (results summarized in Table II) have been performed at the National Research Council (CNR) building in Pisa, Italy. The indoor experimental setup is depicted in the upper side of Figure 8 (reported in the Appendix due to space limitations), with the following configurations. Each test has been performed with the same conditions for 15 times.

- *Experiment 1 (E1)*. In this experiment, U1 moves on a linear pattern while U2 stands still. The two users are physically separated by a wall, as can be seen in Figure 8. The experiment has been performed with U1 moving at two different speeds, namely 6 km/h and 15 km/h, to evaluate the effectiveness of LVS with different walking speeds.
- *Experiment 2 (E2)*. Both U1 and U2 are moving on straight and parallel linear patterns but in opposite directions. As shown in Figure 8, this experiment is performed with several obstacles between U1 and U2. The two users move on parallel trajectories which are 19,2 meters far. The presence of the obstacles and the moving speed caused the authentication protocol to fail 9 times on 15 for the slow speed experiments, and 11 times on 15 for the fast speed experiments.
- *Experiment 3 (E3)*. Same configuration as E2, but with the users moving in the same direction. The experiment has been performed with the users moving

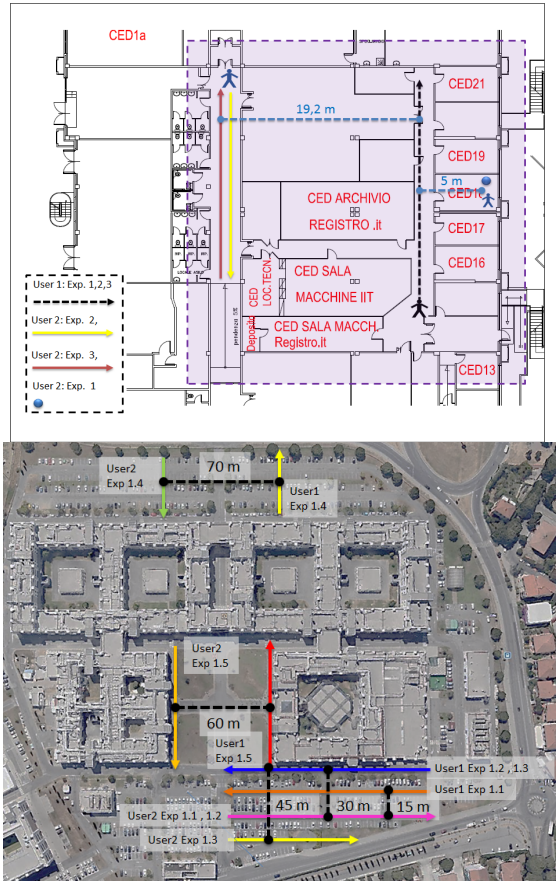

Fig. 8: Indoor and outdoor experimental setups.

at the same speed.

Table II concludes that if users are in the same room or in nearby rooms, it is almost guaranteed the mutual verification will be successful. However, walls and interference caused by other electronic devices may affect the mutual verification, as E2 shows. The improvement in E3 is due to the reduction of the variance of the perceived signal strength between U1 and U2.

The outdoor experiments (summarized in Table III) have been performed in the premises of the National Research Council (CNR) of Pisa, Italy (lower side of Figure 8). Outdoor experiments differ from indoor ones for different distances and obstacles.

In these experiments, U1 and U2 are moving on parallel linear trajectories in opposite directions. Overall, we performed 5 sets of outdoor experiments, with different distances and type of obstacles. As for indoor experiments, each test has been performed 15 times at two different speeds of the users, which are 6 km/h and 15 km/h. Namely, Experiments 1.1, 1.2, and 1.3 (lower side of Figure 8) have been performed with the users moving in different aisles of a parking lot. Each aisle is separated by two rows of cars. Details on the distances and on the obstacles between the users are summarized in Table III. Experiment 1.4 has been performed at the distance of 70 meters in a condition of partial line of sight. Experiment 1.5 has been performed in the central plaza of the CNR area, with the users being in direct line of sight. From this set of experiments, we derive that in absence of walls, with the users in line of sight, the verification is performed always correctly if they are within a range of 60 meters. Experimental results conclude that the LVS approach is viable in both outdoor and indoor environments, given the mutual authentication occurs almost every time in every considered experimental scenario.

|         | Energy ($\mu$Ah) | CI (90%) | Power (mW) | CI (90%) | Lifetime (h) | CI (90%) |
|---------|------------------|----------|------------|----------|--------------|----------|
| LTE     | 5783.20          | ± 524.39 | 399.950    | ± 35.83  | 25.55        | ± 2.22   |
| LTE + LVS | 5810.55        | ± 427.56 | 400.15     | ± 27.69  | 25.42        | ± 2.15   |
| WiFi + LTE | 7346.11       | ± 1103.2 | 480.490    | ± 72.06  | 24.56        | ± 6.19   |

TABLE I: Energy consumed by LTE only vs. LTE + LVS vs. WiFi + LTE.

| Exp. | Distance | 6 km/h | 15 km/h |
|------|----------|--------|---------|
| 1    | 5 m      | 15/15  | 15/15   |
| 2    | 19.2 m   | 6/15   | 4/15    |
| 3    | 19.2 m   | 15/15  | 14/15   |

TABLE II: Details of Indoor Experiments.

| Exp. | Dist. | Obstacles  | 6 km/h | 15 km/h |
|------|-------|------------|--------|---------|
| 1.1  | 15 m  | 2 car lanes | 15/15 | 15/15   |
| 1.2  | 30 m  | 3 car lanes | 15/15 | 15/15   |
| 1.3  | 45 m  | 4 car lanes | 15/15 | 15/15   |
| 1.4  | 70 m  | Partial LoS | 9/15  | 7/15    |
| 1.5  | 60 m  | LoS        | 15/15  | 15/15   |

TABLE III: Details of the outdoor experiments.

### A. Energy consumption evaluation

In order to calculate with high degree of precision the energy efficiency of the LVS location verification algorithm, we set up an experimental testbed as depicted in Figure 9. In particular, we used the *Power Monitor* device [16] to acquire instantaneous power and current consumption of the smartphone. The tool provides a interface by which we could also obtain an estimation of the estimated battery lifetime of the smartphone according to the current energy consumption rate. The measurements are then averaged over a predefined period of time and repeated over different experiments.

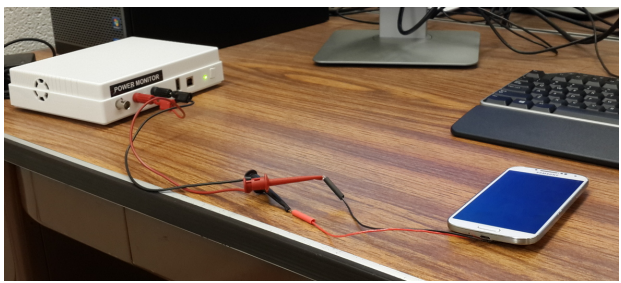

Fig. 9: Experimental setup to calculate energy consumption.

The following experiments have been performed. First, we evaluated the energy consumption of the smartphone while running the sensing app and LTE, with the WiFi interface and LVS localization mechanism turned off. Next, we calculated the energy consumption of the smartphone when the sensing app is running, LTE is active and the LVS localization mechanism is active, which means, the WiFi interface is turned on for $T_{vt}$ = 30 seconds for neighbor discovery and then turned off. Finally, we evaluated the energy consumption of the smartphone when the sensing app is running, WiFi is active 100% of the time and LTE is active. Table I summarizes the energy consumed, the average power consumption and the expected residual lifetime, supposing a battery of 2600 mAh (i.e., the one equipped on Samsung Galaxy S4). The experiments have been conducted by monitoring the energy consumption for 200 seconds and then averaging over different repetitions. In all experiments, the screen was turned off and no other apps were running on the phone; in particular, no app was generating any downstream or upstream traffic, and WiFi was not connected to any network.

Table I remarks the significant difference in energy consumption between WiFi and LTE. In particular, Table I shows a difference of almost 1600 $\mu$Ah between WiFi and LTE. This was expected, given that WiFi consumes huge amount of energy even when not connected to a network [23]. However, the most important result indicated by Table I is that LVS has almost no impact on the energy consumption of the smartphone, as WiFi is turned on only for a short period of time (30 seconds), and then remains turned off most of the time.

### B. Simulations results

In this section, we evaluate through simulation experiments the performance of LVS in terms of resilience from attackers and efficiency. To simulate a realistic environment, we modeled the sensing area as a single location area large 4 square kilometers (size of a small city or city block). As far as user mobility is concerned, we assumed users move about the location area following the Truncated Lévi Walk (TLW) mobility model [24], which has been shown to best represent the mobility of humans [25]. Due to space limitations, we refer the reader to [25] for additional insights.

For the sake of simplicity, we modeled the WiFi range of the smartphones devices as circles centered on the user with radius 50 meters. As default system parameters, we chose as reputation parameters $\Delta_b = 0.25$, $\Delta_d = 0.6$, and $\Delta_u = 0.15$. The setup time $T_{sw}$ has been set to 7 seconds according to the experimental evaluation of Section II, while the validation round time $T_{vr}$ has been set to 15s. The validation epoch threshold $M$ and $\theta$ have been respectively set to 0.9 and 0.8, while the $q$ parameter has been set to 2 in all experiments. The confidence intervals are set to 95%. For the sake of graphical clarity, the confidence intervals are not shown when less than 1% of the average. In the following, we will refer to as "users" the participants not faking their position, and to "attackers" as participants who fake their position and implement the LSA attack described in Section 2. For the sake of simplicity, and without losing in generality, we also assumed that users remain active inside the same location area for the whole duration of a validation epoch.

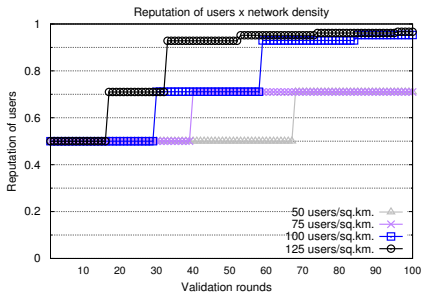 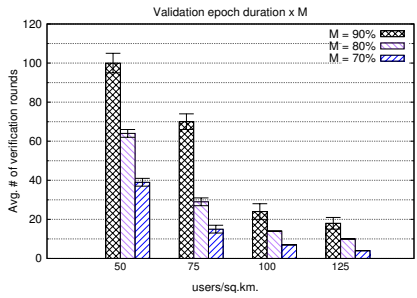 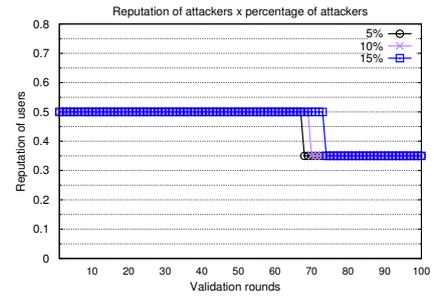

Fig. 10: Reputation of users (users density).

Fig. 11: Average time of validation epochs.

Fig. 12: Reputation of attackers (50 users/sq.km).

First, we evaluate the impact of the user density on the users' reputation and the efficiency of LVS. Specifically, Figure 10 shows the average reputation opinion of users as function of user density, supposing no attackers are present in the location area. As expected, from Figure 10 we observe that to greater user density corresponds faster increase of user reputation level over time, which is given by the faster termination of each validation epoch of LVS. This is further validated by Figure 11, which depicts the average duration of the validation epochs of LVS as function of users density and the $M$ parameter. Recall that in LVS, a validation epoch ends when $M$ percent of users have their position validated by at least $q$ users.

Figure 11 concludes the validation epoch duration decreases as $M$ decreases and the users density increases, which is coherent to what depicted in Figure 10. This means that increasing the $M$ parameter allows a more complete validation of users' location, to the expense of longer validation epochs. This trade-off should be met by the administrator of the PS application by considering the average users density and the desired level of security.

As expected, it also turns out (results not shown here for the sake of space) that the validation epoch duration increases as the WiFi range decreases. This is reasonable, since more time connectivity will be available to users at each validation round. However, a shorter WiFi range implies that the validation of users' position will be finer-grained. Therefore, the WiFi range parameter may be set by the administrator of the PS application depending on the desired trade-off between precision and efficiency of LVS.

| Users x sq.km. | % of MHSs | C.I. |
|---|---|---|
| 50 | 14.5 | 2.51 |
| 75 | 17.33 | 3.12 |
| 100 | 21.75 | 3.56 |
| 125 | 24.25 | 4.11 |

TABLE IV: Number of users selected as MHSs.

Finally, to further validate the scalability of the LVS approach, Table reports the percentage of the users selected as MHSs in function of the user density. Table concludes that the percentage of users selected as MHS is significantly less than the total number of users, even when the density become relatively high.

*1) Resilience from attackers:* Let us now evaluate the resilience of LVS to attackers with Figures 12 and 13, which show the average reputation of all the attackers in function of the percentage of attackers in the system. Specifically, the attack has been simulated by setting the position of all attackers outside the location area, and by making them advertise a random position inside the location area to the PS platform. We recall that we do not consider in this analysis colluding attackers. Figures 12 and 13 conclude that the attackers' reputation decreases faster when the users density is higher, due to the shorter duration of validation rounds. Therefore, LVS is able to detect faster users not advertising their real position when the users density is higher. However, as anticipated earlier, note that LVS does **not** increase the reputation level of attackers in any circumstance, given the location of the attackers will never be validated by any MHS. Also, note that the reputation of attackers never reaches the $\theta$ threshold necessary to accept their reports inside the PS system. Therefore, we conclude LVS is able to exclude unreliable reports from the PS system and therefore **protects** the PS system from the location-spoofing attack defined in Section II, without compromising the functionality of the PS application. We would like to point out that the security parameters of the reputation algorithm, as well as the validation round time $T_{vt}$, may be tuned by the administrator of the PS application according to the desired trade-off between efficiency and security.

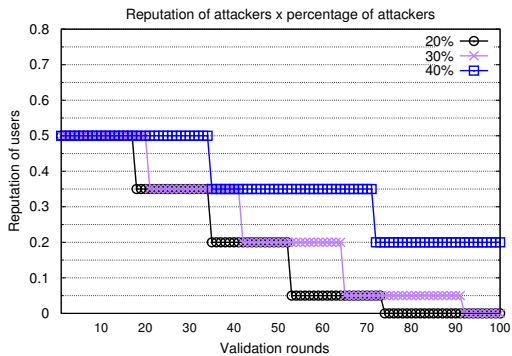

Fig. 13: Reputation of attackers (125 users/sq.km.).

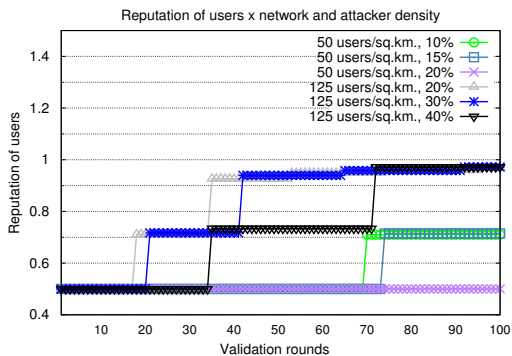

Fig. 14: Reputation of users (network and attacker density).

To gain further insights on the impact of the attackers on the reputation of users, Figure 14 show the reputation of users as function of the percentage of attackers and the users density. Figure 14 shows that when the density of users is relatively low (50 users/sq.km.), more validation epochs are needed to increase the reputation of users. Simply enough, this is due to the fact that in this case the users density on the location area becomes very low, and therefore LVS takes additional time to validate the users' locations. However, Figure 14 also remarks that when the density of users is relatively high (125 users/sq.km.), LVS is able to tolerate a very high percentage of attackers (40%) without hindering the reputation of users. This is because in this case LVS will still maintain enough users to validate each user's location and therefore will tolerate a higher number of attackers.

## V. RELATED WORK

In this section, we survey existing work related to the localization of smartphones and users, and highlight the novel contributions brought by this paper.

Given location-spoofing software like `Fake Locator` is able to hijack both GPS and GSM location services, approaches such the one presented in [26] are prone to the LSA attack and therefore not suitable to validate user location in PS systems. In addition, the user location obtained through GSM cell triangulation is known by the telephone service providers only, and may not be shared with external parties due to privacy issues. Conversely, the LVS framework does not require any piece of information that cannot be retrieved on smartphones, which is essential for easy deployment.

Existing WiFi-based solutions [27], [28] were specifically designed for indoor environment only, and are therefore not applicable to large-scale outdoor PS systems. Instead, LVS leverages a technique that is valid for both indoor *and* outdoor PS systems. Although [27], [28] and similar solutions yield a greater accuracy than LVS, we point out here that LVS is not aimed at calculating the precise location of users. Instead, the goal of LVS is to *verify* the user location provided by other localization services and thus solve LSA attacks. Finally, approaches based on ambient-based fingerprints [15], [29] are not suitable in PS scenarios in which users are not able to observe the same phenomenon (e.g., users located in different floors/rooms of a building). The proposed LVS framework, instead, is *independent* of the collected data type and relies only on WiFi to verify user position. A trust algorithm for distributed environments, similar to the one proposed in this work has been exploited in [30] to verify attribute values in usage control systems with faulty Attribute Managers, and in [20] where it has been exploited to manage a collaborative framework for Android malware analysis.

## VI. CONCLUSIONS

In this paper, we have proposed LVS, a location validation system which verifies user location in participatory sensing (PS) systems and solves the proposed location-spoofing attack (LSA). First, we have proposed LVS, which authenticates user location in a distributed and scalable way through the use of the mobile WiFi hotspot capability of modern smartphones. Furthermore, we have introduced the formalism of the Chains of Sight, which are used to implement an algorithm to tackle collusion-based attacks. We have also proposed a reputation-based system based on LVS which rules out reports coming from users spoofing their location. Finally, we have tested the proposed approach in indoor and outdoor testbeds, measured its energy consumption, and shown its effectiveness against the LSA attack through simulations. Results conclude that LVS is energy-efficient, applicable in almost every practical PS scenarios, and effectively solves LSA-based attacks.

TABLE OF SYMBOLS

| Symbol | Meaning | Placement |
|---|---|---|
| $u_i, u_j, \ldots$ | Generic user | III |
| $A_i, A_j, \ldots$ | Generic Location Area | III |
| $A_k^t$ | Location Area of user $k$ at time $t$ | III-C |
| $T_r$ | Validation round duration | III-C |
| $T_e$ | Validation epoch duration | III-C |
| $T_{sw}$ | WiFi setup time of MHS | III-C |
| $T_{vt}$ | Validation time of MHS | III-C |
| $N_i^{t_j}$ | Users present in $A_i$ at time $t_j$ | III-C |
| $D_i^{t_j}$ | Users declared in $A_i$ at time $t_j$ | III-C |
| $\psi_{max}$ | Max duration of validation epoch | III-F |
| $\Omega_i$ | Area knowledge of user $u_i$ | III-F |

TABLE V: Table of Symbols